\documentclass[preprint]{article}
\usepackage{graphicx}
\usepackage{dcolumn}
\usepackage{bm}
\usepackage{amsmath}
\usepackage{amssymb}
\usepackage{multirow}
\usepackage{caption}
\usepackage{epstopdf}
\usepackage{hyperref}

\begin{document}
{\setlength{\oddsidemargin}{1.2in}
\setlength{\evensidemargin}{1.2in} } \baselineskip 0.55cm
\begin{center}
{\LARGE {\bf Impact of Buchdahl metric potential on thin-shell gravastar framework in de Rham-Gabadadze-Tolley like massive gravity }}
\end{center}
\date{\today}
\begin{center}
  Meghanil Sinha*, S. Surendra Singh \\
Department of Mathematics, National Institute of Technology Manipur,\\
Imphal-795004,India\\
Email:{ meghanil1729@gmail.com, ssuren.mu@gmail.com}\\
 \end{center}
 
\textbf{Abstract}: This paper presents a study on gravitational vacuum stars (gravastars) with an isotropic matter distribution in de Rham-Gabadadze-Tolley (dRGT) massive gravity incorporating Buchdahl metric function. Here we have conducted an analysis on thin-shell singularity-free gravastar configuration. The study demonstrates the viability of gravastars as alternatives to black holes (BHs) in this massive gravity. Our research yields singularity free analytical solutions for gravastar interior and without event horizon. Our discussion focuses on the properties of the thin shell of ultra-relativistic stiff fluid viz., length, energy, entropy and the massive gravity's impact on these physical properties. The junction conditions have been carefully examined and study of surface redshift analysis implies regularity of the model. Our investigation also includes energy condition analysis supporting the thin shell formation. Thus our solutions eliminate singularities and information paradoxes and the implications of these solutions are seemingly noteworthy and exhibit physical desirable properties.\\

\textbf{Keywords}: Gravastar, black holes, massive gravity.\\

\section{Introduction}\label{sec1}

\hspace{0.5cm}General relativity (GR) paved the way for modern physics which has driven advancements in cosmology as well as astrophysics. Einstein's work established the foundation of matter-geometry interplay which led him to describe gravity as spacetime curvature. The endpoint of stellar collapse is a matter of ongoing discussion in astrophysics analogous to the Universe's primordial state in cosmology. Einstein's GR theory has been extensively verified in accordance with the observational outcomes in either fields. In scenarios with extreme energetic phenomena, such as the Universe's early stages and stellar collapse's endpoint, due to the confinement of significant energy in very small volume, causing the energy density to diverge leading to the formation of the singularities. Here the GR's framework cease to apply allowing for quantum corrections to mitigate the singular behavior \cite{GR1}. Pertaining to the terminal state of stellar collapse, the influence of quantum effects sheds new light to the well-known BH end states such as the phenomenon Hawking radiation from the BH event horizons. However, this approach does not eliminate the singularity in the particular outcomes in case of the field equations \cite{GR2,GR3}.\\
The Schwarzschild singularity at $ R=2GM $ ( for $ c=1 $ ) does not correspond to a physical singularity. This can be regularized via co-ordinate choice. But the central singularity at the origin ($ r=0 $) is a physical one and cannot be removed. Mazur and Mottola's research led to the proposal of the novel gravastar concept in 2001, as a potential substitute for the BH, which was further refined and expanded in 2004 \cite{GR4,GR5}. With quantum considerations it was proposed that the horizon could be regarded as a surface of critical phenomenon associated with a gravitational phase shift for a matter distribution with an equation of state (EoS) given by $ p=-\rho $. The negative pressure contributes to a force that is repulsive \cite{GR7,GR8,GR9}. In the light of the fact that the horizon is characterized by a transition of phase, Mazur and Mottola explored the quantum fluctuations that govern the behavior of the energy-momentum tensor's components at the horizon and appears significant enough to produce an EoS of the form  $ p=\rho $. This type of EoS is close to instability and the most minimal extreme allowed and resulting the interior to form a gravitational Bose-Einstein condensate (BEC). Zeldovich's stiff fluid replaces the critical surface \cite{GR10,GR11}. The exterior, the third region is marked by an absence of pressure and entropy density. Further research investigated the stability of this star building on Mazur and Mottola's foundation by accounting for the physical attributes and EoS \cite{R1}. A key idea was that gravastars without shells must possess anisotropic pressures \cite{R2}. The stability of gravastars under generic perturbation schemes was demonstrated \cite{R3}. Quasi-normal modes provide a way to differentiate gravastars from BHs of the same mass. Gravastar solutions were explored in the framework of Born-Infeld phantom theory \cite{R4}. Stability analysis of static, spherically symmetric gravastars was also carried out and it was revealed that anisotropic gravitational vacuum stars with continuous pressure exhibit radial stability \cite{R5}.\\
The acceleration of the Universe at late times has been validated through various independent observations \cite{S2,S3}. Dark energy (DE) characterized by negative pressure was introduced to account for the Universe's accelerating nature. Lambda, or the cosmological constant represents a basic form of DE. The $\Lambda$CDM model ( combining Lambda and cold dark matter ) provides a successful description of the Universe's current era. The model encounters theoretical difficulties, a mismatch with the theoretical predicted values. Additionally it also struggles with the ratio of CDM and that of ordinary matter to DE in the present epoch. Given its immense success, GR remains as the top choice for mathematically describing various cosmological phenomena. GR's forecast of gravitational waves was realized with LIGO/VIRGO's observational evidence \cite{S9}. While powerful, GR is unlikely the last word on gravitational theory. It faces challenges in accounting for inflation and the Universe's accelerating expansion \cite{S10}. This led to the development of various modified gravity theories to address GR's limitations. A significant body of work focuses on gravastar models within modified gravities \cite{RT,RSIGMAT,Rastall,QT,RLmT,Braneworld,Mimetic}. Research on gravastar also extends to higher dimensional and $ (2+1) $ dimensional spacetime frameworks \cite{Dimension1,Dimension}. Studies have investigated the role of electromagnetic charge in various gravastar models \cite{Charge2,Charge3}. \\
In the context of GR, gravitons are viewed as massless spin-2 particles \cite{M2,M3}. The theory of massive gravity modified GR to include gravitons with small mass. Pauli's 1939 work introduced linear massive gravity \cite{M4}. Non-linear extensions of this theory were subsequently studied \cite{M4,M5}. But ghost instability is a frequent consequence of non-linearity. As a result new ghost-free massive gravity theories emerged in non-linear contexts. One notable ghost free massive gravity is de Rham-Gabadadze-Tolley (dRGT) like massive gravity.  dRGT effectively resolves the ghost instability, including the Boulware-Deser ghosts, in non-linear configurations \cite{M9,M10,M11}. The Vainshtein mechanism masks the extra degree of freedom linked to the matter source via non-linear effects. This mechanism allows this massive gravity to satisfy various tests. A massive graviton results in a Yukawa-type gravitational potential, with the range determined by the graviton mass. The change results in the deviation of gravitational forces in varied range. A massive graviton's mass might effectively replace the cosmological constant in the role of driving Universe acceleration \cite{M12}. More intriguingly, offers a possible framework for understanding late-time acceleration without DE. The theory was constructed by introducing massive graviton terms in the gravitational action. BH solutions in dRGT massive gravity has been explored in some studies \cite{M15,M16}. Determining consistency with observations in cosmology and astrophysics would be a notable achievement for this modified massive gravity theory. dRGT massive gravity's implications for compact structures have been studied \cite{M21,M22}. Inspired by the previous works, in this study we present complete gravastar solutions for different regions with corresponding EoS in dRGT massive gravity utilizing the Buchdahl metric potential. To ensure stability, we have studied surface redshift along with the junction conditions for the thin shell formation.\\
In the present manuscript, section (\ref{sec1}) provides a concise overview of the gravastars and massive gravity while section (\ref{sec2}) explores the EFE for gravastar configurations within dRGT massive gravity incorporating the Buchdahl metric potential. Section  (\ref{sec3}) elaborates on the non-singular nature of the interior solution of the stellar model and section  (\ref{sec4}) is dedicated to the shell region. The vacuum spacetime with Schwarzschild solution is covered in section s (\ref{sec5}). We investigate the junction conditions between the interior and exterior regions in section  (\ref{sec6}). Analysis of gravastar stability using the surface redshift is presented in section (\ref{sec7}). Section (\ref{sec8}) and (\ref{sec9}) cover physical features and energy conditions respectively. Section (\ref{sec10}) gives the concluding observations.\\

\section{Construction of field equation in dRGT massive gravity}\label{sec2}

\hspace{0.5cm}The action for Einstein gravity with nonlinear interaction terms as graviton mass is as:
\begin{equation}\label{1}
\hat{S} = \frac{1}{16 \pi G}\int[\check{R}+M_{g}^{2}V(g,\Psi^{\alpha})]\sqrt{-g}d^{4}x + \int L_{m}\sqrt{-g}d^{4}x.
\end{equation}\\
Here $G$ symbolize the Universal gravitational constant and $g$ the determinant of the metric tensor $g_{mn}$ respectively. $\check{R}$ denotes the Ricci scalar and $ M_{g} $ the graviton mass. Matter lagrangian is represented by $ L_{m} $. The graviton potential due to the non-zero graviton mass is\\
\begin{equation}\label{2}
V(g,\Psi^{\alpha}) = V_{2} + aV_{3} + bV_{4}
\end{equation}\\
where $a$ and $b$ represent dimensionless parameters. $V_{2},V_{3}$ and $V_{4}$ are determined by\\
\begin{eqnarray}\label{3}
V_{2} &=& [K]^{2} - [K^{2}]
\nonumber \\
V_{3} &=& [K]^{3} - 3[K][K^{2}] +2[K^{3}]
\nonumber \\
V_{4} &=& [K]^{4} - 63[K]^{2}[K^{2}] + 8[K][K^{3}] + 3[K^{2}]^{2} - 6[K^{4}]
\end{eqnarray}\\
\begin{equation}\label{4}
K_{j}^{i} = \delta_{j}^{i} - \sqrt{g^{ih}\acute{F}_{\alpha \beta}\partial_{h}\Psi^{\alpha}\partial_{h}\Psi^{\beta}}
\end{equation}\\
$[\hspace{0.7cm}]$ signifies the trace of rank-two tensor $ K_{j}^{i} $.
\begin{equation}\label{5}
[K] = K_{i}^{i}, \hspace{1cm} [K^{\gamma}] = (K^{\gamma})_{i}^{i}.
\end{equation}\\
$\acute{F}_{\alpha \beta}$ is the fiducial metric acting as a reference metric and $ \Psi^{\alpha} $ serve as Stuckelberg fields. The unitary guage $ \Psi^{\alpha} = x^{i}\delta_{i}^{\alpha} $ is used as\\
\begin{equation}\label{6}
\sqrt{g^{ih}\acute{F}_{\alpha \beta}\partial_{h}\Psi^{\alpha}\partial_{h}\Psi^{\beta}} = \sqrt{g^{ih}\acute{F}_{hj}}.
\end{equation}\\
In accordance with the previous work, we opt for the simple fiducial metric \cite{M15,M21} as\\
\begin{equation}\label{7}
\acute{F}_{\alpha \beta} = diag(0, 0, U^{2}, U^{2}\sin^{2}\theta)
\end{equation}\\
where $U$ denotes a positive valued constant. Variation of equation (\ref{1}) w.r.t $ g_{\mu\lambda} $ gives\\
\begin{equation}\label{8}
\check{R}_{\mu\lambda} - \frac{1}{2}\check{R}g_{\mu\lambda} + M_{g}^{2}\eta_{\mu\lambda} = \frac{8 \pi G}{c^{4}} T_{\mu\lambda}
\end{equation}\\
$G$ and $c$ are set to be $1$ in this paper. The graviton's energy-momentum tensor is given as \\
\begin{eqnarray}\label{9}
\eta_{\mu\lambda} &=& \frac{1}{\sqrt{-g}}\frac{\delta \sqrt{-g}V}{\delta g^{\mu\lambda}}
\nonumber \\
&& = K_{\mu\lambda} - Kg_{\mu\lambda} - (3a+1)[K_{\mu\lambda}^{2} - KK_{\mu\lambda} - \frac{[K]^{2} - [K^{2}]}{2} g_{\mu\lambda}] +
\nonumber \\
&& 3(a+4b)[K_{\mu\lambda}^{3} - KK_{\mu\lambda}^{2} + \frac{[K]^{2} - [K^{2}]}{2} K_{\mu\lambda} - \frac{1}{6} g_{\mu\lambda}([K]^{3}-3[K][K^{2}]+2[K^{3}])].
\end{eqnarray}\\
Due to the fiducial metric's form, $ O(K^{4}) $ terms do not appear. The dRGT massive gravity tensor is expressed as \\
\begin{equation}\label{10}
T_{\mu\lambda}^{grav} = -\frac{M_{g}^{2}}{8\pi}\eta_{\mu\lambda} = -(\rho^{grav} + p^{grav})u_{\mu}u_{\lambda} + pg_{\mu\lambda}.
\end{equation}\\
For normal matter, the energy momentum tensor is given as\\
\begin{equation}\label{11}
T_{\mu\lambda} = -\frac{2}{\sqrt{-g}}\frac{\partial \sqrt{-g}L_{m}}{\partial g^{\mu\lambda}} = (\rho + p)u_{\mu}u_{\lambda} + pg_{\mu\lambda}.
\end{equation}\\
$M_{g}^{2}\eta_{\mu\lambda}$'s components are specified \cite{M25} as\\
\begin{eqnarray}\label{12}
\rho^{grav} &=& \frac{M_{g}^{2}}{8 \pi}\eta_{t}^{t} 
\nonumber \\
&& = - \frac{M_{g}^{2}}{8 \pi} \Big(\frac{(3a+1)(3r-U)(r-U)}{r^{2}} + \frac{3(a+4b)(r-U)^{2}}{r^{2}} + \frac{3r-2U}{r}\Big)
\nonumber \\
&& = \frac{1}{8\pi r^{2}}\Big(\Pi r^{2} - 2\Xi r - \Sigma \Big)
\end{eqnarray}\\
\begin{eqnarray}\label{13}
p^{grav} &=& -\frac{M_{g}^{2}}{8 \pi}\eta_{r}^{r} 
\nonumber \\
&& = \frac{M_{g}^{2}}{8 \pi} \Big(\frac{(3a+1)(3r-U)(r-U)}{r^{2}} + \frac{3(a+4b)(r-U)^{2}}{r^{2}} + \frac{3r-2U}{r}\Big)
\nonumber \\
&& = -\frac{1}{8\pi r^{2}}\Big(\Pi r^{2} - 2\Xi r - \Sigma \Big)
\end{eqnarray}\\
and \\
\begin{eqnarray}\label{14}
p^{grav} &=& -\frac{M_{g}^{2}}{8 \pi}\eta_{\theta,\phi}^{\theta,\phi} 
\nonumber \\
&& = \frac{M_{g}^{2}}{8 \pi} \Big(\frac{(3a+1)(2U-3r)}{r} + \frac{3(a+4b)(U-r)}{r} + \frac{U-3r}{r}\Big)
\nonumber \\
&& = -\frac{1}{8\pi r}\Big(\Pi r - 2\Xi \Big)
\end{eqnarray}\\
where\\
\begin{eqnarray}
\Pi &=& -3M_{g}^{2}(1+A+B)
\nonumber \\
\Xi &=& UM_{g}^{2}(1+2A+3B)
\nonumber \\
\Sigma &=& U^{2}M_{g}^{2}(A+3B)
\end{eqnarray}\\
and here we have chosen 
\begin{eqnarray}\label{16}
A &=& 3a+1
\nonumber \\
B &=& a+4b
\end{eqnarray}\\
The relations above allow us to express $U, A$ and $B$ using $\Pi, \Xi$ and $\Sigma$ as\\
\begin{equation}\label{17}
A = \frac{\Xi\sqrt{\Xi^{2} + (M_{g}^{2} + \Pi)\Sigma} - (2M_{g}^{2} + \Pi)\Sigma - \Xi^{2}}{M_{g}^{2} \Sigma}
\end{equation}\\
\begin{equation}\label{18}
B = \frac{2\Pi}{3M_{g}^{2}} - \frac{\Xi\sqrt{\Xi^{2} + (M_{g}^{2} + \Pi)\Sigma} - M_{g}^{2}\Sigma - \Xi^{2}}{M_{g}^{2} \Sigma}
\end{equation}\\
\begin{equation}\label{19}
U = \frac{\sqrt{\Xi^{2} + (M_{g}^{2} + \Pi)\Sigma} + \Xi}{M_{g}^{2} + \Pi}.
\end{equation}\\
In the limit $ \Sigma\rightarrow 0 $, $U, A$ and $B$ are finite and thus we have \cite{M21}\\
\begin{eqnarray}\label{20}
U &=& \frac{2\Xi}{M_{g}^{2} + \Pi}
\nonumber \\
A &=& -3B = -\frac{3}{2} - \frac{\Pi}{2M_{g}^{2}}.
\end{eqnarray}\\
The metric is given as\\
\begin{equation}\label{21}
ds^2=-e^{2m}dt^2+e^{2n}dr^2+r^2(d\theta^2+\sin^2\theta d\phi^2).
\end{equation}\\
Using equation (\ref{21}), the field equations from equation (\ref{8}) are read as\\ 
\begin{equation}\label{22}
e^{-2n}\Big(\frac{2n'}{r} - \frac{1}{r^{2}} \Big) + \frac{1}{r^{2}} = 8\pi\rho + \frac{1}{r^{2}}\big(\Pi r^{2} - 2\Xi r\big)
\end{equation}\\
\begin{equation}\label{23}
e^{-2n}\Big(\frac{2m'}{r} + \frac{1}{r^{2}} \Big) - \frac{1}{r^{2}} = 8\pi p - \frac{1}{r^{2}}\big(\Pi r^{2} - 2\Xi r\big)
\end{equation}\\
\begin{equation}\label{24}
e^{-2n}\big(m'' + m'^{2} - m'n' - \frac{n'-m'}{r} \big) - \frac{1}{r^{2}} = 8\pi p - \frac{1}{r}\big(\Pi r - \Xi \big).
\end{equation}\\  
We have adopted here the Buchdahl form for the metric potential $ e^{2m(r)} $ as \cite{B1}\\
\begin{equation}\label{25}
e^{2m} = \frac{P(Qr^{2}+1)}{P+Qr^{2}} \hspace{1cm} 0<P<1 
\end{equation}\\
where $Q$ represents a free parameter characterising the metric function. The metric function as per our requirement exhibit finite and well behaved nature at the origin\\
\begin{equation}\label{26}
e^{2m(0)} = 1 \hspace{1cm} \frac{\partial}{\partial r}e^{2m(r)}\Big|_{r=0} = 0.
\end{equation}\\
Substituting the above conditions into the field equations (\ref{22}-\ref{24}) we find\\
\begin{equation}\label{27}
e^{-2n}\Big(\frac{2n'}{r} - \frac{1}{r^{2}} \Big) + \frac{1}{r^{2}} = 8\pi\rho + \frac{1}{r^{2}}\big(\Pi r^{2} - 2\Xi r\big)
\end{equation}\\
\begin{equation}\label{28}
e^{-2n}\Big(-\frac{2PQ(1+Qr^{2})}{(P+Qr^{2})^{2}} + \frac{2PQ}{P+Qr^{2}} + \frac{1}{r^{2}} \Big) - \frac{1}{r^{2}} = 8\pi p - \frac{1}{r^{2}}\big(\Pi r^{2} - 2\Xi r\big)
\end{equation}\\
\begin{eqnarray}\label{29}
e^{-2n}\Big(\frac{8PQ^{2}r^{2}(1+Qr^{2})}{(P+Qr^{2})^{3}} - \frac{8PQ^{2}r^{2}}{(P+Qr^{2})^{2}} - \frac{2PQ(1+Qr^{2})}{(P+Qr^{2})^{2}}
\nonumber \\
 + \frac{2PQ}{P+Qr^{2}} + \frac{4(P-1)^{2}P^{2}Q^{2}r^{2}}{(P+Qr^{2})^{4}} + n'\big(\frac{2PQ(1+Qr^{2})}{(P+Qr^{2})^{2}} - \frac{2PQr}{P+Qr^{2}} \big)
\nonumber \\
 -\frac{n'}{r} - \frac{2PQ(1+Qr^{2})}{(P+Qr^{2})^{2}} + \frac{2PQ}{P+Qr^{2}} \Big) = 8\pi p - \frac{1}{r}\big(\Pi r - \Xi \big).
\end{eqnarray}\\
The energy momentum tensor conservation equation yields\\
\begin{equation}\label{30}
\frac{dp}{dr} + (\rho + p )\frac{dm}{dr} = 0 
\end{equation}\\

\section{Interior region}\label{sec3}

\hspace{0.5cm}The gravastar's interior EoS is characterised by $ p = -\rho $. This EoS produces a repulsive force here, and a gravitational BEC emerges after the phase transition at the horizon ( replaced with a shell ) directing outward, resisting further collapse. With the EoS, equation (\ref{30}) yields $ p = -\rho = \rho_{e} $. $ \rho_{e} $ being the constant interior density implies constant pressure throughout. With this, the field equation (\ref{27}) provides the other metric potential in the form of\\
\begin{equation}\label{31}
e^{-2n} =  1- \frac{8\pi \rho_{e} r^{2}}{3} - \frac{\Pi r^{2}}{3} +\Xi r + \frac{Y}{r}.
\end{equation}\\
For regularity at origin, we take $Y=0$ which leaves us with\\
\begin{equation}\label{32}
e^{-2n} =  1- \frac{8\pi \rho_{e} r^{2}}{3} - \frac{\Pi r^{2}}{3} + \Xi r.
\end{equation}\\ 
\begin{figure}[h!]
\centering
\includegraphics[scale=0.5]{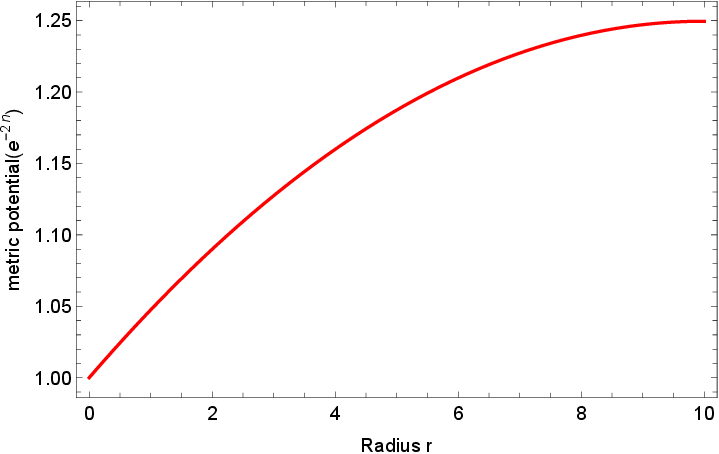}
\caption{Plot of the metric potential $e^{-2n}$ in the interior versus radial co-ordinate for $\Pi=0.005$, $\Xi=0.05$, $\rho_{e}=0.0001$ }\label{1}
\end{figure}\\
Figure (\ref{1}) presents the metric potential $ e^{-2n} $'s behavior in the interior, where it is evident that the interior solutions are singularity free. As a result, the central singularity problem of the classical black hole (CBH) can be resolved. For the gravastar's interior, the active gravitational mass can be determined using\\
\begin{equation}\label{33}
\bar{M} = \int_{0}^{r_{1}=R} 4 \pi r^{2}\rho dr =  \frac{4\pi R^{3} \rho_{e}}{3}.
\end{equation}\\
\begin{figure}[h!]
\centering
\includegraphics[scale=0.5]{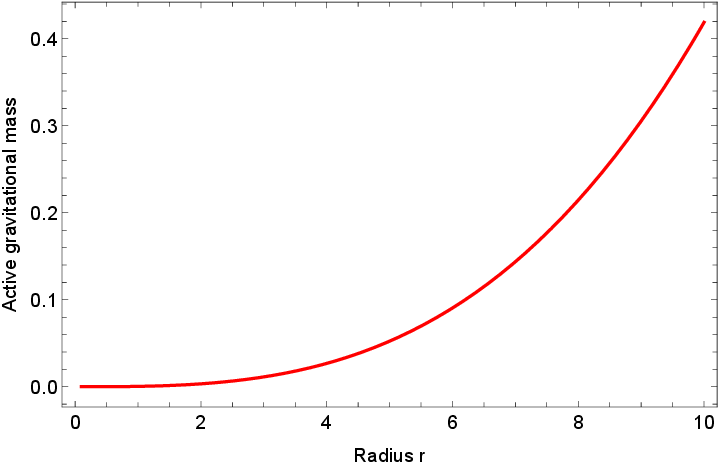}
\caption{Plot of active gravitational mass in the interior versus radial co-ordinate for $\rho_{e}=0.0001$ }\label{2}
\end{figure}\\
As depicted in figure (\ref{2}), the $ \bar{M} $ exhibits a well behaved nature due to its non-singularity and smoothness.\\

\section{Shell region}\label{sec4}

\hspace{0.5cm}The intermediate zone of the gravitational vacuum star is represented by this region. This shell is modeled as an ultra-relativistic fluid adhering to a specific EoS $ p=\rho $. It possesses extremely small thickness, such that $ e^{-2n}\ll 1 $ resulting in a shell thickness between $0$ and $1$. These suppositions facilitate simpler calculations. Zeldovich's fluid formation has been analyzed here \cite{Z1}. This fluid concept has been extensively explored in different fields \cite{Z2,Z3}. With this thin shell approximation and the series expansion, the field equations (\ref{27}-\ref{29}) give\\
\begin{equation}\label{34}
e^{-2n} = 2\Xi r - \Pi r^{2} + \log[r] - X
\end{equation}\\
\begin{figure}[h!]
\centering
\includegraphics[scale=0.5]{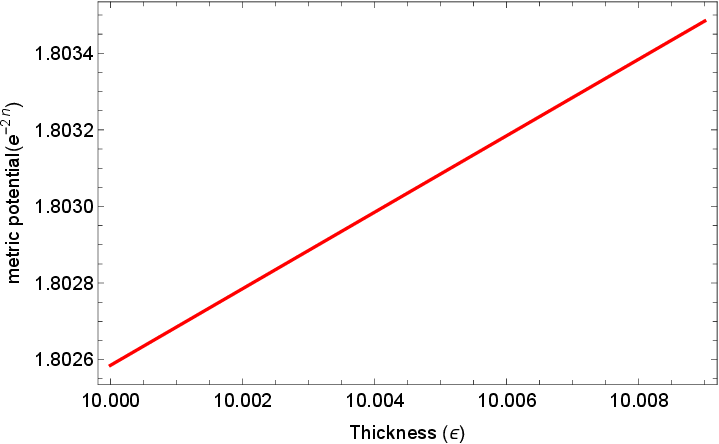}
\caption{Plot of the metric potential $e^{-2n}$ in the shell versus shell thickness for $\Pi=0.005$, $\Xi=0.05$ }\label{3}
\end{figure}\\
where $X$ represents the constant of integration. Figure (\ref{3}) highlights the variation of $ e^{-2n(r)} $ in the shell region. It is smooth, well-behaved and devoid of any type of singularities. Equations (\ref{25}), (\ref{30}) and the thin shell EoS yield\\
\begin{equation}\label{35}
\rho = p = We^{\frac{(P-1)P}{P+Qr^{2}}}
\end{equation}\\
\begin{figure}[h!]
\centering
\includegraphics[scale=0.5]{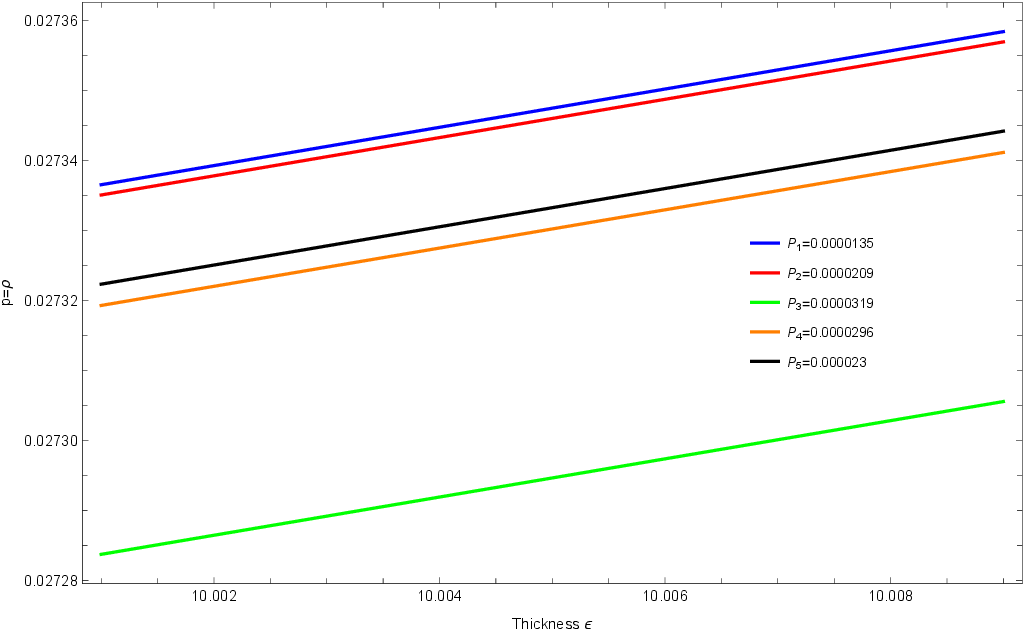}
\caption{Plot of the metric potential $e^{-2n}$ in the shell versus shell thickness for  $ (i) P = 0.0000135, Q = -7.455 \times 10^{-8} km ^{2} $,  $ (ii) P = 0.0000209, Q = -1.1446 \times 10^{-7} km^{2} $, $ (iii) P = 0.0000319, Q = -1.5664 \times 10^{-7} km^{2} $, $ (iv) P = 0.0000296, Q = -1.4545 \times 10^{-7} km^{2} $, $ (v) P = 0.000023, Q = -1.154 \times 10^{-7} km^{2} $ }\label{4}
\end{figure}\\
with $W$ being the integration constant. Figure (\ref{4}) depicts an increase in density from the interior boundary to the exterior \cite{RSIGMAT,Rastall}. Thus the exterior junction is more compact form compared to the interior.\\

\section{Exterior region}\label{sec5}

\hspace{0.5cm}The exterior of the gravastar satisfies the EoS $ p = \rho = 0 $. This implies that the exterior is a vacuum. The Schwarzschild metric provides the line element for the outside region provided by,\\
\begin{equation}\label{36}
ds^{2} = \Big(1- \frac{2M}{r}\Big)dt^{2} - \frac{1}{\Big(1- \frac{2M}{r}\Big)}dr^{2} - r^{2}d\sigma^{2}
\end{equation}\\
where $ d\sigma^{2} = (d\theta^{2} + \sin ^{\theta}d\phi^{2}) $. $ M $ signifies the object's total mass.\\

\section{Junction condition and the EoS parameter}\label{sec6}

\hspace{0.5cm}The gravastar's structure includes three regions: the interior (I), the exterior (III) and the intermediate shell (II). Hence, it is fundamental to the gravastar's framework. It results in a manifold that is geodetically complete, with matter confined to the surface shell $ (r = R) $. This results in a single manifold describing the gravastar. The junction condition necessitates a smooth matching for the regions (I) and (III). Using the Dramois-Israel condition, we will calculate the surface stresses at the interface \cite{Re1,Re2}. The Lanczos equation provides the intrinsic surface stress-energy tensor $ S_{lm} $ in the form \cite{Re4,Re5}:\\
\begin{equation}\label{37}
S_{m}^{l} = -\frac{1}{8\pi}\Big(\kappa_{m}^{l} - \delta_{m}^{l}\kappa_{k}^{k} \Big).
\end{equation}\\
The second fundamental form's discontinuity is described by\\
\begin{equation}\label{38}
\kappa_{lm} = k_{lm}^{+} - k_{lm}^{-}. 
\end{equation}\\
This allows us to represent the second fundamental form as\\
\begin{equation}\label{39}
k_{lm}^{\pm} = -n_{i}^{\pm}\Big[ \frac{\partial^{2}x_{i}}{\partial\chi^{l}\partial\chi^{m}} + \Gamma_{ab}^{i}\frac{\partial x^{a}\partial x^{b}}{\partial\chi^{l}\partial\chi^{m}} \Big] \Big|_{S}
\end{equation}\\
with $ \chi ^{l} $ being the intrinsic co-ordinates on the shell. $ n_{i}^{\pm} $ represents the unit normals to $ S $(the surface), given a spherically symmetric static metric
\begin{equation}\label{40}
ds^{2} = u(r)dt^{2} - u(r)^{-1}dr^{2} - r^{2}(d\theta^{2} + \sin ^{\theta}d\phi^{2}).
\end{equation}\\
$ n_{i}^{\pm} $ is expressed as\\
\begin{equation}\label{41}
n_{i}^{\pm} = \pm \Big| g^{ab} \frac{\partial u}{\partial x^{a}} \frac{\partial u}{\partial x^{b}} \Big|^{-\frac{1}{2}} \frac{\partial u}{\partial x^{i}}.
\end{equation}\\
where $ n^{\alpha}n_{\alpha} = 1 $. Via the Lanczos equation, we find the surface stress energy tensor to be $ S_{lm} = diag[D, -P, -P, -P] $. In this context, $D$ denotes the surface energy density and $P$ the surface pressure. The parameters are defined by\\
\begin{equation}\label{42}
D = -\frac{1}{4 \pi R}\Big[\sqrt{u}\Big]_{-}^{+}
\end{equation}\\
\begin{equation}\label{43}
P = -\frac{D}{2} + \frac{1}{16 \pi} \Big[\frac{u'}{\sqrt{u}}\Big]_{-}^{+}.
\end{equation}\\
By combining equations (\ref{36}) and (\ref{32}) along with the preceding equations we get $D$ and $P$ as
\begin{equation}\label{44}
D = -\frac{1}{4 \pi R}\Big[ \sqrt{1- \frac{2M}{R}} - \sqrt[4]{1- \frac{8\pi \rho_{e} R^{2}}{3} - \frac{\Pi R^{2}}{3} + \Xi R} \Big]
\end{equation}\\
\begin{eqnarray}\label{45}
P &=& \frac{1}{16 \pi} \times \Big(\frac{2M}{\sqrt{1- \frac{2M}{R}} R^{2}} +
\nonumber \\
&& \frac{-3\Xi + 2(\Pi + 8 \pi \rho_{e})R}{2 \times 3^{\frac{1}{4}} ( 3 + 3\Xi R - \Pi R^{2} - 8\pi \rho_{e} R^{2} )^{\frac{3}{4}}} + 2 \big(\sqrt{1- \frac{2M}{R}}-(1+\Xi R - \Pi \frac{R^{2}}{3} - \frac{8\pi \rho_{e} R^{2}}{3})^\frac{1}{4}\big) \Big).
\end{eqnarray}\\
\begin{figure}[h!]
\centering
\includegraphics[scale=0.5]{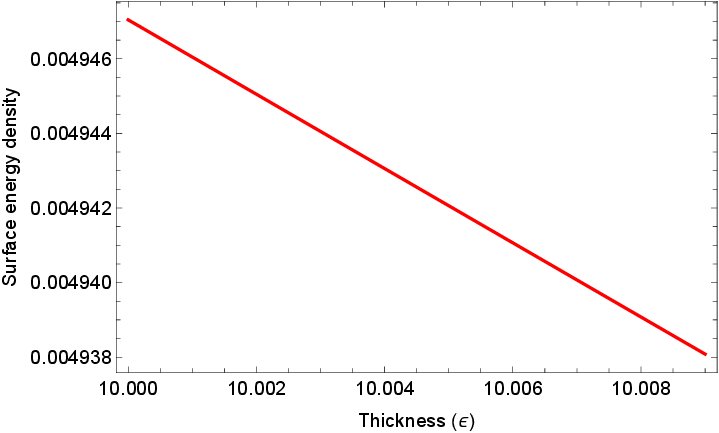}
\caption{Plot of surface energy density versus shell thickness for $\Pi=0.005$, $\Xi=0.05$, $\rho_{e}=0.0001$ and $ M = 3.75 M_{\bigodot} $ }\label{5}
\end{figure}\\
\begin{figure}[h!]
\centering
\includegraphics[scale=0.5]{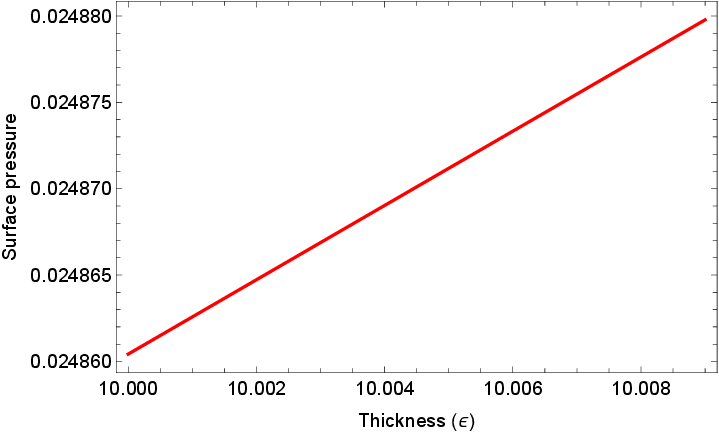}
\caption{Plot of surface pressure versus shell thickness for $\Pi=0.005$, $\Xi=0.05$, $\rho_{e}=0.0001$ and $ M = 3.75 M_{\bigodot} $ }\label{6}
\end{figure}\\
Figures (\ref{5}) and (\ref{6}) visualize how the surface energy density and surface pressure vary w.r.t thickness of the shell. We note that both the parameters are positive throughout the entire region. This shows that the null energy condition is met, enabling the formation of the shell, as demonstrated in \cite{RSIGMAT,Mimetic}. Figure (\ref{5}) shows a decline in $D$ as the thickness increases \cite{RSIGMAT,QT}. With surface energy density equation (\ref{44}), the mass of the thin shell is computable as\\
\begin{eqnarray}\label{46}
Mass_{Shell} &=& 4 \pi R^{2} D
\nonumber \\
&& = -R\Big( \sqrt{1- \frac{2M}{R}} - \big(1- \frac{8\pi \rho_{e} R^{2}}{3} - \frac{\Pi R^{2}}{3} + \Xi R\big)^{\frac{1}{4}} \Big).
\end{eqnarray}\\
The above equation enables us to calculate the gravastar mass as a function of the thin shell mass as\\
\begin{eqnarray}\label{47}
M &=& \frac{1}{6R} \times 
\nonumber \\
&& \Big(-3Mass_{Shell}^{2} + 3R^{2} + 2 \times 3^{\frac{1}{4}}Mass_{Shell}R( 3 + 3\Xi R - \Pi R^{2} - 8\pi \rho_{e} R^{2} )^{\frac{1}{4}} 
\nonumber \\
&& - \sqrt{3}R^{2}\sqrt{ 3 + 3\Xi R - \Pi R^{2} - 8\pi \rho_{e} R^{2} } \Big).
\end{eqnarray}
The EoS parameter has the form as\\
\begin{equation}\label{48}
P = \varpi (R) D.
\end{equation}\\
Equations (\ref{44}) and (\ref{45}) yield the EoS parameter in the explicit form as\\
\begin{equation}\label{49}
\varpi (R) = -\frac{1}{2} + \frac{R\Big(\frac{2M}{\sqrt{1- \frac{2M}{R}} R^{2}} + \frac{-3\Xi + 2(\Pi + 8 \pi \rho_{e})R}{2 \times 3^{\frac{1}{4}} ( 3 + 3\Xi R - \Pi R^{2} - 8\pi \rho_{e} R^{2} )^{\frac{3}{4}}}\Big)}{4\Big(\sqrt{1- \frac{2M}{R}} - \big(1- \frac{8\pi \rho_{e} R^{2}}{3} - \frac{\Pi R^{2}}{3} + \Xi R \big)^{\frac{1}{4}} \Big)}.
\end{equation}\\
\begin{figure}[h!]
\centering
\includegraphics[scale=0.5]{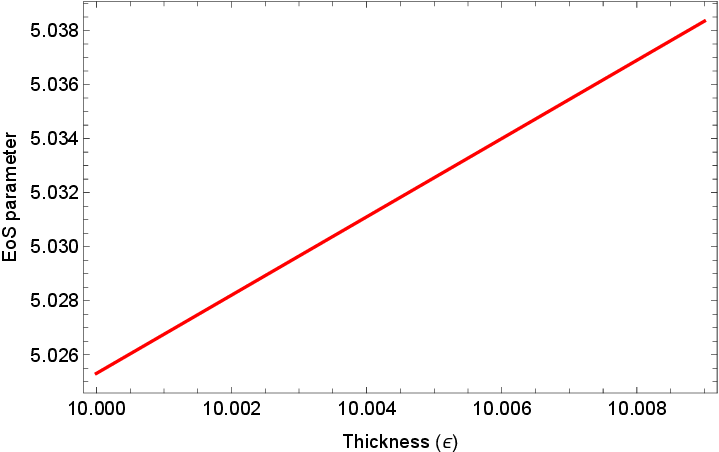}
\caption{Plot of EoS parameter versus shell thickness for $\Pi=0.005$, $\Xi=0.05$, $\rho_{e}=0.0001$ and $ M = 3.75 M_{\bigodot} $ }\label{7}
\end{figure}\\
The variation of $\varpi (R) $ in figure (\ref{7}) reveals a decrease in its nature with the shell thickness.\\

\section{Stability analysis}\label{sec7}

Here, we analyze the surface redshift to assess the stability of the thin-shell gravastars in dRGT massive gravity. With the form of $ \verb"Z"_{s} = \frac{\Delta \lambda}{\lambda_{e}} = \frac{\lambda_{o}}{\lambda_{e}} $, we have the means to calculate the surface redshift, where $ \lambda_{e} $ is the detected wavelength ( by the observer ) and $ \lambda_{o} $ is the emitted wavelength ( from the source ). A surface redshift limit of 2 has been suggested for stable isotropic fluid distributions \cite{ST1,ST2}. For anisotropic fluids, its' limit has been suggested as high upto 3.84 \cite{ST3}. In the absence of cosmological constant, it has been demonstrated that $ \verb"Z"_{s}  \leq 2 $ for isotropy case \cite{ST4}. For anisotropic stars with cosmological constant, the condition $ \verb"Z"_{s}  \leq 5 $ has been shown. We have calculated $ \verb"Z_{s}" $ from the formula as:
\begin{equation}\label{50}
\verb"Z"_{s}  = -1+\frac{1}{\sqrt{g_{tt}}} = \frac{1}{e^{m}} - 1.
\end{equation}\\
Thus we arrive at surface redshift as\\
\begin{equation}\label{51}
\verb"Z"_{s}  = \sqrt{\frac{P+Qr^{2}}{P(Qr^{2}+1)}} - 1.
\end{equation}
\begin{figure}[h!]
\centering
\includegraphics[scale=0.5]{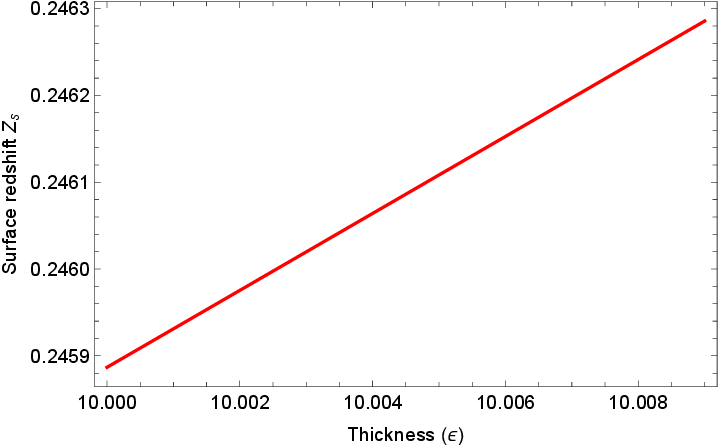}
\caption{Plot of surface redshift  versus shell thickness for  $ (i) P = 0.0000135, Q = -7.455 \times 10^{-8} km ^{2} $ }\label{8}
\end{figure}\\
Figure (\ref{8}) provides a graphical analysis of  $ \verb"Z_{s}" $. The figure shows that it is well within the stability bounds \cite{QT}. This suggests that the gravastar model is stable and physically relevant in this massive gravity.\\

\section{Some physical properties}\label{sec8}

\hspace{0.5cm}The goal of this section is to examine the impact of the massive parameter along with the Buchdahl metric function on the physical properties of the gravastars in this massive gravity. The proper length, energy and entropy inside the thin shell are the subjects of our investigation in this section. To visualize our results we facilitate the graphical representations.\\

\subsection{Proper length}

\hspace{0.5cm}The thin shell approximation assumes the boundaries at $R$ and $R+\epsilon (\epsilon \ll 1)$. Accordingly, the proper length of the shell is thereby calculated as\\
\begin{equation}\label{52}
\mathcal{L} = \int_{R}^{R+\epsilon} \sqrt{e^{2n}} dr.
\end{equation}\\
With the substitution from equation (\ref{34}) it leads to\\
\begin{equation}\label{53}
\mathcal{L} = \Big(\frac{4 \Xi r - \Pi r^{2} + 2\log [r] - 2X\log [r]}{4\sqrt{2}(\Xi)^{\frac{3}{2}}}\Big)_{R}^{R+\epsilon}.
\end{equation}\\
\begin{figure}[h!]
\centering
\includegraphics[scale=0.5]{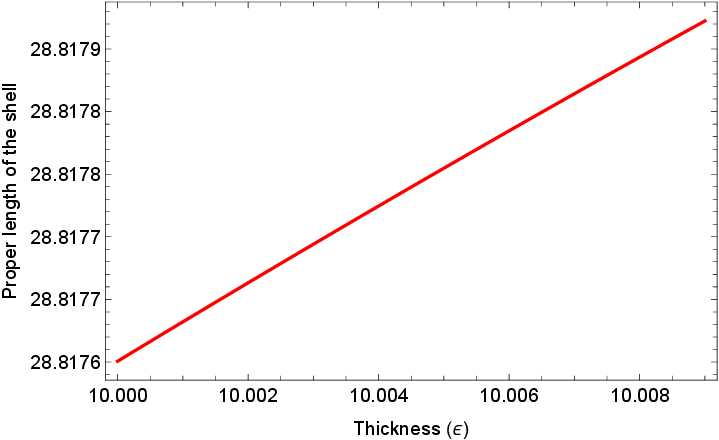}
\caption{Plot of proper shell length  versus shell thickness for  $\Pi=0.005$, $\Xi=0.05$ }\label{9}
\end{figure}\\
The shell length-thickness relationship is shown graphically in figure (\ref{9}). There is a clear linear relationship, with the length growing as the thickness increases \cite{Braneworld}.\\

\subsection{Energy}

\hspace{0.5cm}We define the energy within the shell as \\
\begin{equation}\label{54}
\varepsilon = \int_{R}^{R+\epsilon} 4 \pi r^{2} \rho dr.
\end{equation}\\
Using the equation (\ref{35}) we get\\
\begin{equation}\label{55}
\varepsilon = 4 \pi W \Big[\frac{(P-1)Pr}{Q} + \frac{r^{3}}{3} - \frac{(P-1)P^{\frac{3}{2}}\arctan [\frac{\sqrt{Q}r}{\sqrt{P}}]}{Q^{\frac{3}{2}}}\Big]_{R}^{R+\epsilon}.
\end{equation}\\
\begin{figure}[h!]
\centering
\includegraphics[scale=0.5]{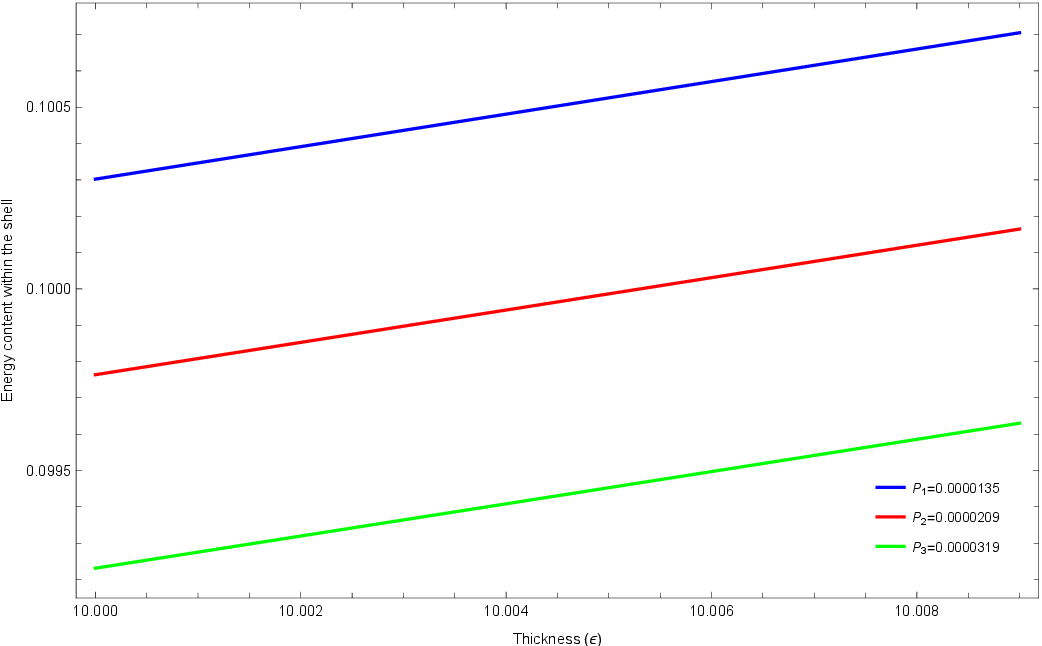}
\caption{Plot of shell energy  versus shell thickness for   $ (i) P = 0.0000135, Q = -7.455 \times 10^{-8} km ^{2} $,  $ (ii) P = 0.0000209, Q = -1.1446 \times 10^{-7} km^{2} $, $ (iii) P = 0.0000319, Q = -1.5664 \times 10^{-7} km^{2} $ }\label{10}
\end{figure}\\
The change in the shell energy is shown in figure (\ref{10}). The plot reveals a rise in the energy with thickness. Energy fluctuations mirror the matter density fluctuations. The condition of increasing energy is thus fulfilled \cite{RT,Rastall}.\\

\subsection{Entropy}

\hspace{0.5cm}Mazur and Mottola's approach implies zero entropy density, consistent with a single condensate phase \cite{GR4,GR5}. Within the shell, entropy takes the form of\\
\begin{equation}\label{56}
\mathfrak{S} = \int_{R}^{R+\epsilon} 4 \pi r^{2} s(r) \sqrt{e^{2n}} dr
\end{equation}\\
where $s$ is the local entropy density, and\\
\begin{equation}\label{57}
s(r) = \frac{b^{2}k_{B}^{2}T(r)}{4 \pi \hbar^{2}} = b\frac{k_{B}}{\hbar}\sqrt{\frac{p}{2\pi}}.
\end{equation}\\
Here $b$ denotes a dimensionless constant in this context. With $ k_{B} = \hbar = 1 $, we get\\
\begin{equation}\label{59}
s(r) = b\sqrt{\frac{p}{2\pi}}. 
\end{equation}\\
The shell's entropy is thereby obtained by the substitution from equations (\ref{34}) and (\ref{35}) as\\
\begin{eqnarray}\label{60}
\mathfrak{S} &=& -\frac{1}{8 \Xi Q^{2}}\times
\nonumber \\
&& \Big[Q^{2}r^{2} - \Pi Q^{2}r^{4} + Q^{2}(-\Pi (P-1)P + 2XQ - 2Q\log[r])
\nonumber \\
&& + (P-1)P(\Pi P +XQ - Q\log[r])\log[1+\frac{\sqrt{Q}r}{\sqrt{P}}]+ (P-1)P(\Pi P +XQ
\nonumber \\
&& - Q\log[r])\log[1+\frac{\sqrt{Q}r}{P^{\frac{3}{2}}}] - (P-1)PQ PolyLog[2,\frac{\sqrt{Q}r}{\sqrt{P}}]
\nonumber \\
&& - (P-1)PQ PolyLog[2,\frac{\sqrt{Q}r}{P^{\frac{3}{2}}}] \Big]_{R}^{R+\epsilon}
\end{eqnarray}\\
\begin{figure}[h!]
\centering
\includegraphics[scale=0.5]{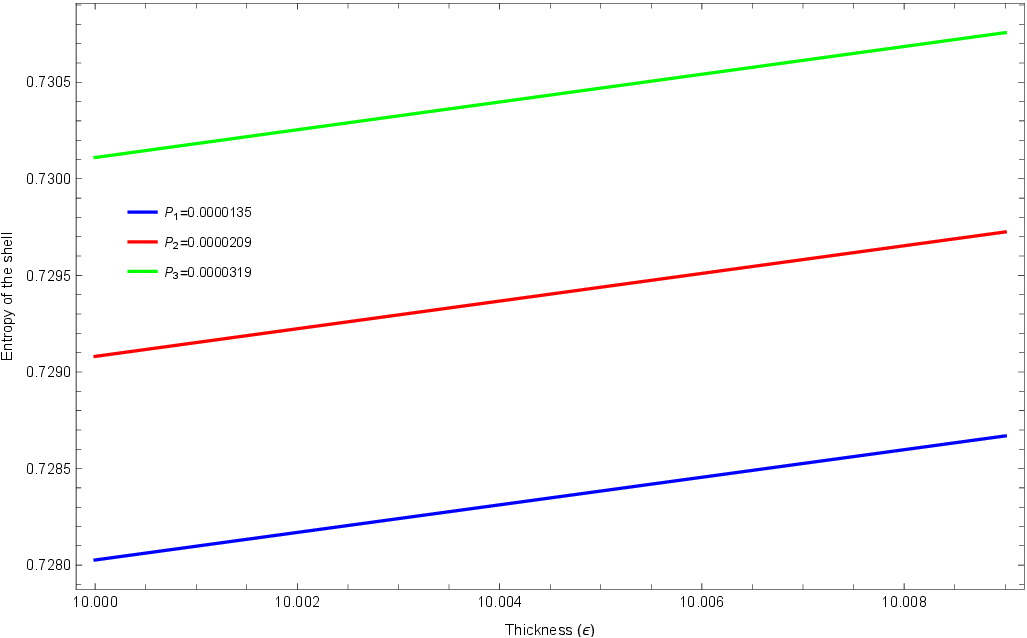}
\caption{Plot of shell entropy  versus shell thickness for   $ (i) P = 0.0000135, Q = -7.455 \times 10^{-8} km ^{2} $,  $ (ii) P = 0.0000209, Q = -1.1446 \times 10^{-7} km^{2} $, $ (iii) P = 0.0000319, Q = -1.5664 \times 10^{-7} km^{2} $, $\Pi=0.005$, $\Xi=0.05$}\label{11}
\end{figure}\\
where $PolyLog$ is the poly-logarithm function $ Li_{s}= z $. The entropy variation of the shell with thickness is plotted in figure (\ref{11}). It indicates an increasing shell entropy. A stable gravastar configuration requires a maximum entropy on the surface \cite{QT}.\\

\section{Energy conditions}\label{sec9}

To be physically viable, stellar structures must adhere to the energy conditions. The commonly known conditions are:\\
$1.$ NEC : $ D + P > 0 $\\
$2.$ WEC : $ D > 0 $ and $ D + P > 0 $\\
$3.$ SEC : $ D + P > 0 $ and $ D + 3P > 0 $\\
$4.$ DEC : $ D > 0 $ and $ D \pm P > 0 $.\\
\begin{figure}[h!]
\centering
\includegraphics[scale=0.5]{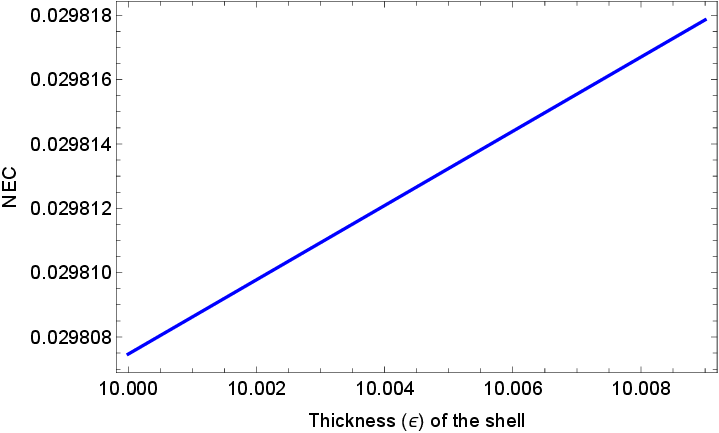}
\caption{Plot of surface NEC versus shell thickness for $\Pi=0.005$, $\Xi=0.05$, $\rho_{e}=0.0001$ and $ M = 3.75 M_{\bigodot} $ }\label{12}
\end{figure}\\
\begin{figure}[h!]
\centering
\includegraphics[scale=0.5]{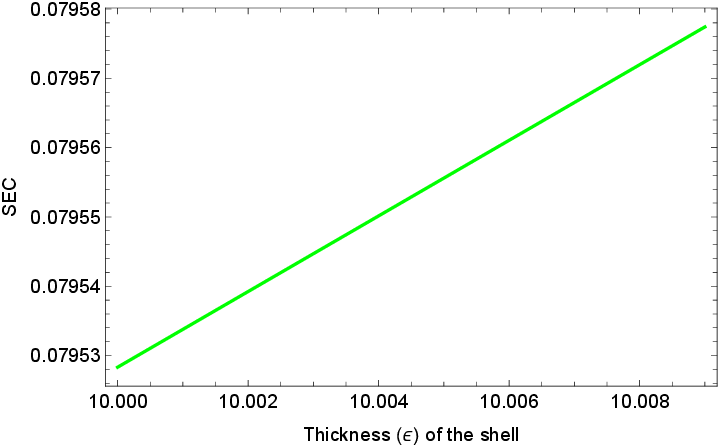}
\caption{Plot of surface SEC versus shell thickness for $\Pi=0.005$, $\Xi=0.05$, $\rho_{e}=0.0001$ and $ M = 3.75 M_{\bigodot} $ }\label{13}
\end{figure}\\
\begin{figure}[h!]
\centering
\includegraphics[scale=0.5]{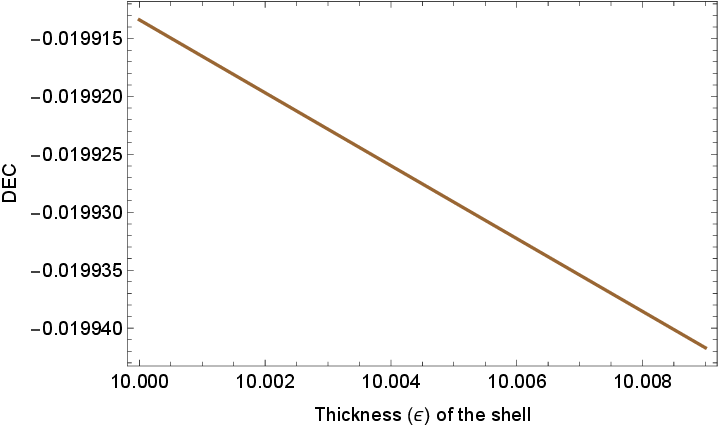}
\caption{Plot of surface DEC versus shell thickness for $\Pi=0.005$, $\Xi=0.05$, $\rho_{e}=0.0001$ and $ M = 3.75 M_{\bigodot} $ }\label{14}
\end{figure}\\
Satisfying the energy conditions ensures the model's physical validity.  We are testing the null energy requirement to determine if exotic matter resides in the shell. Our analysis of figures (\ref{12}-\ref{14}) shows NEC and SEC satisfaction but DEC violation for the thin shell. As a result, energy flux appears negative \cite{Mimetic}.\\

\section{Conclusion}\label{sec10}

\hspace{0.5cm}In this research we have developed a gravastar solution in de Rham-Gabadadze-Tolley like massive gravity focusing on the Buchdahl metric potential. It can be viewed as a realistic alternative to the BHs. The Buchdahl metric potential's non-singular and regular nature makes it suitable for gravastar's geometry. The three gravastar regions have been presented here. Our analysis provides regular solutions for the interior geometry. The thin shell's characteristics are investigated for particular values of the parameters demonstrating the gravastar's physical validity.\\
With the dark energy EoS, the conservation equation implies a constant matter density in the inner region. The uniform density produces a outward force, stabilizing the structure. Our solution for the metric function $e^{2n}$ (figure \ref{1}) in the interior geometry is well behaved and finite at the core. The gravitational mass is found to be positive throughout here (figure \ref{2}). As a result, this ensures the absence of central singularity. The field equations are solved for the metric function $e^{2n}$ in the shell, assuming ultra-relativistic matter. The Schwarzschild metric has been used to describe the exterior geometry. The surface energy density and surface pressure are obtained using the Darmois-Israel matching requirements. Our analysis covers the key aspects of the EoS parameter in figure (\ref{7}). Surface redshift analysis confirms our model's stability. In this case the redshift parameter is found to be within 1(stability range).\\
Important features such as shell length, energy and entropy have been studied in details in section (\ref{8}) by taking the help of the figures (\ref{9}), (\ref{10}) and (\ref{11}). Together, these parameters point to a stable model. Our investigation also shows that WEC, NEC and SEC are satisfied but DEC is violated for the emergence of the thin shell as illustrated in figures (\ref{12}), (\ref{13}) and (\ref{14}). These findings confirm the physical relevance of this particular Buchdahl metric function in dRGT massive gravity. Hence, this massive gravity yields stable and realistic models of massive compact stars under the given metric. Exploring gravastar regions with diverse metric functions in dRGT massive gravity and beyond would be rather interesting with a promising avenue.\\

\end{document}